# Maxwell's Equations in Media and Their Solution[*]

Tao Zhang

College of Nuclear Science and Technology, Beijing Normal University, Beijing 100875

taozhang@bnu.edu.cn

**Abstract** Magnetic field magnetization, polarization and induced magnetization are analyzed. It is shown that the existing Faraday's law of electromagnetic induction is not reasonable in media. A modified Faraday's law of electromagnetic induction in media is put forward and used to revise existing Maxwell's Equations in media. A solution of the revised Maxwell's Equations is presented.

## 1. Introduction

Maxwell's electromagnetic theory leads to the discovery of electromagnetic waves (EWs). Maxwell's Equations in vacuum have been verified with many experiments. Besides EWs in vacuum, the interaction between media and EWs is also an important subject. Recent related researches include millimeter wave imaging and metamaterial etc. The contradictions between the existing theory and some new experimental results indicate that the interaction between media and EWs is a complicated subject. In this paper, propagation of the EWs in the media is studied, and some viewpoints are presented. For simplicity, "media" in this paper means insulation, homogeneous, isotropic and infinite media, and the media are in quiescent state. "Electromagnetic waves (EWs)" means only the refraction parts of EWs, and does not include the lost parts of EWs due to reflection, absorption and scattering etc.

## 2. Recently-proposed viewpoints about refraction mechanism of EWs

### 2.1 The electron cloud conductors

Ref. [1] indicates that each electron cloud in the molecules of the media can be seen as a tiny conductor. The electron-cloud conductors are very small, so that the electric fields and the magnetic fields of the EWs can penetrate them. When the alternating magnetic field $B$ of the EWs exists in the electron-cloud conductor, $dB/dt$ induces an electromotance in the electron-cloud conductor, and there may be an induced current on the electron-cloud conductor. The induced current is actually a statistical result of the electron's motion, and is an additive directional motion superposed on the original motion of the electron in its electron cloud. The magnetization by the induced current existing in the electron-cloud conductor is called the "induced magnetization"[2]. For distinguishing, the traditional magnetization is called the "magnetic field magnetization".

---

[*] Beijing Science Technology New Star Program (Grant No. 952870400). The Chinese version of this paper can be found on http://www.paper.edu.cn

Hence, in addition to the magnetic field magnetization *M* and the polarization *P*, there may be the induced magnetization $M_F$ in the media under the EWs. *M* caused by the magnetic field. $M_F$ caused by variation of the magnetic field, and has a different phase with the magnetic field. $M_F$ is related to *dB*/*dt*, not *B* [2]. *M* is related to *B*, not *dB*/*dt*.

## 2.2 Refraction mechanism of EWs in media

Refraction of the EW in the media is caused by the induced magnetization, not by the polarization and the magnetic field magnetization, because the energy of the induced magnetization is not a part of the energy of the EW, while the energies of the magnetic field magnetization and the polarization are parts of the energy of the EW itself [1]. This can be explained by the phase difference $\Delta\phi$ between the magnetic field of the EW and the magnetic field of the induced magnetization (i.e. the magnetic field of the induced current on the electron-cloud conductor). $\Delta\phi =\pi$, see Fig.1. $\Delta\phi =\pi/2$ was presented in our previous articles, but now we believe $\Delta\phi =\pi$.

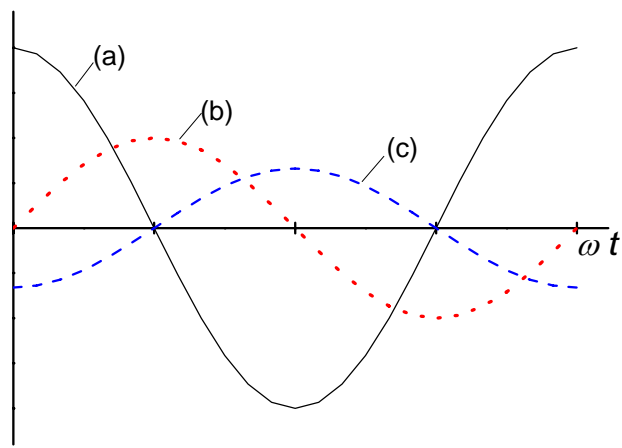

Figure 1 Phases of some variables. (a) *B*, i.e. the magnetic field of the EW; (b) the induced electromotance in the electron-cloud conductor by *dB*/*dt*; (c) the magnetic field of the induced current on the electron-cloud conductor.

Since the magnetic field of the induced magnetization is in different phase with *B*, it is not involved in the mutual transformation process of electromagnetic fields of the EW, and the energy of the induced magnetization should not be regarded as a component of the energy of the EW, although it comes from the EW. Ref. [1] calls the energy of the magnetic field of the induced magnetization the "refractive energy", and believes that during propagation of the EW in the media the EW exchanges the refractive energies back and forth with the electrons in the media (the electrons have main contribution to the exchanges, and the contributions of protons etc. can be omitted). It is the energy exchanges that cause the EWs' slowing down in the media (EWs refraction). This mechanism will be discussed further in Section 4.1.

According to above refraction mechanism and the principle of energy conservation, a calculation method of the refractive index is deduced [1]. The refractive indices of helium, neon, argon, air and the alcohol solutions have been calculated with the method. The results are listed as in Tables 1 and 2.

Table 1 The calculated and the measured refractive indices of some substances

| substances | helium | neon | argon | air |
|---|---|---|---|---|
| calculated refractive indices ($n'$-1) | $3.46 \times 10^{-5}$ [3] | $7.0 \times 10^{-5}$ [4] | $2.82 \times 10^{-4}$ [5] | $2.915 \times 10^{-4}$ [1] |
| measured refractive indices ($n$-1) | $3.5 \times 10^{-5}$ [6] | $6.9 \times 10^{-5}$ [6] | $2.83 \times 10^{-4}$ [6] | $2.92 \times 10^{-4}$ [7] |

Table 2 The calculated and the measured refractive indices of the alcohol solutions

| alcohol content/wt% | 4.00 | 13.00 | 17.00 | 19.00 | 28.00 | 34.00 | 35.00 |
|---|---|---|---|---|---|---|---|
| calculated refractive indices $n'$ [1] | 1.33497 | 1.34112 | 1.34397 | 1.34536 | 1.35102 | 1.35402 | 1.35446 |
| measured refractive indices $n$ [8] | 1.33497 | 1.34109 | 1.34396 | 1.34537 | 1.35109 | 1.35402 | 1.35447 |
| $n-n'$ | 0 | −0.00003 | −0.00001 | 0.00001 | 0.00007 | 0 | 0.00001 |

The calculation results of refractive-index are in good agreement with the measured values in the handbooks.

## 2.3 Quantum characteristics of the refractive energy, and influences of frequency of EWs on refraction ability of media

The refractive energies are holed by the electrons in the molecules. Since the electrons are confined in the molecules, according to the theory of quantum mechanics, the refractive energies must be of quantization. Therefore, as the EW interacts with the electron in the molecules, only if the electromagnetic induction is strong enough, i.e. the energy quantum of the EW is large enough, to make the electron absorb the refractive energy from the EW. This situation is similar to the photoelectric effect [9]. Using above viewpoint to analyze the relationship between frequency of the EWs and the refractive abilities of the molecules, we can anticipate [10]: (1) In the range of light frequency, the electromagnetic induction between light and the electrons in the molecules is strong enough, and all the molecules have enough refractive abilities; (2) As the frequency decreasing, firstly the inner electrons and then the outer electrons in the molecules gradually do not absorb refractive energies, and the molecules lose their refractive abilities gradually; (3) In the range of microwave frequency, the electrons in most molecules do not absorb refractive energies, and the molecules and the media made of them in nature have little refractive abilities.

Researches on millimeter-wave imaging utilizing refractive dielectric lenses attract a lot of attentions at present. This technology has many potential applications, such as in security check, military, medicine and transportation. But its imaging quality is not satisfactory so far. The dielectric lenses are mostly made of polymer materials. Ref. [10] believes that most of the polymer materials have weak refractive abilities to the millimeter waves. That should be one of the reasons for the low quality of the imaging.

## 2.4 Method for improving materials' refractive abilities to millimeter waves

The refractive indices of the traditional artificial materials have not reached the expected values [11]. Ref. [10] believes that the conductors in the traditional artificial materials have weak electromagnetic inductions when they meet the millimeter waves, and that the cage-shaped granule of conductor (CGC) should have a strong electromagnetic induction as it meets the millimeter waves, because CGC includes closed loops of conductor. The millimeter waves can pass through the closed loops and produce strong electromagnetic inductions with CGC. So strong induction currents form on CGC, and that makes CGC possess enough refractive ability to the millimeter waves. The experimental results show that CGC materials have considerable refractive abilities to the millimeter wave, while the polymer materials have little refractive ability to the millimeter wave [10].

In a word, it is the induced magnetization that results in refraction of the EWs in the media. This viewpoint is not consistent with the traditional one. The traditional viewpoint is that the magnetic field magnetization and the polarization result in refraction of the EWs in the media. Since the former opinion is supported by several good verifications of the experimental data in Tables 1 and 2, we have good reason to doubt the rationality of the latter opinion.

## 3. Problem with Maxwell's Equations in media

Ref. [12] believes that the existing Maxwell Equations in media are not quite reasonable. The existing Maxwell's Equations in the media are [13]

$$\nabla \cdot \boldsymbol{E} = 0, \tag{1}$$

$$\nabla \times \boldsymbol{E} = -\frac{\partial \boldsymbol{B}}{\partial t}, \tag{2}$$

$$\nabla \cdot \boldsymbol{B} = 0, \tag{3}$$

$$\nabla \times \boldsymbol{B} = \mu_0 \frac{\partial \boldsymbol{P}}{\partial t} + \mu_0 \nabla \times \boldsymbol{M} + \mu_0 \varepsilon_0 \frac{\partial \boldsymbol{E}}{\partial t}, \tag{4}$$

where $\boldsymbol{E}$ is the electric field intensity, $\boldsymbol{B}$ the magnetic induction, $\boldsymbol{P}$ the polarization and $\boldsymbol{M}$ the magnetic field magnetization in the media. Equation (2) is Faraday's Law of induction. It can be expressed as

$$\oint \boldsymbol{E} \cdot d\boldsymbol{l} = -\iint \frac{\partial \boldsymbol{B}}{\partial t} \cdot d\boldsymbol{S}. \tag{5}$$

Equations (2) and (5) do not include the polarization term and the magnetic-field-magnetization term. Therefore they do not apply to using in the media. The following example [12] is offered to illustrate its failure in the media. Suppose that there are a toroid made of the medium and a varying magnetic field $\boldsymbol{B}_0$ passing through the toroid ($\boldsymbol{B}_0$ is in vacuum), see Fig.2. Let $\boldsymbol{B}_0$ be perpendicular to plane S which is circled by center line $l$ of the toroid, and the symmetry axis of the magnetic field $\boldsymbol{B}_0$ coincide with that of the toroid, as shown in Fig.2. Thus the induced electric field caused by $-\iint \frac{\partial \boldsymbol{B}_0}{\partial t} \cdot d\boldsymbol{S}$ coincides with the center line $l$ of the toroid, and the absolute value of $\boldsymbol{E}$ is the same one everywhere on the center line $l$. The induced electric field caused by $-\iint \frac{\partial \boldsymbol{B}_0}{\partial t} \cdot d\boldsymbol{S}$ makes the medium (the toroid) polarized. Suppose the polarization on the center line $l$ is $\boldsymbol{P}$. $\boldsymbol{P}$ is in the same direction as the induced electric field because of the symmetry. Let $\frac{\partial \boldsymbol{B}_0}{\partial t}$ keep unchanged during a period of time ($\frac{\partial \boldsymbol{B}_0}{\partial t} \neq 0$), then $-\iint \frac{\partial \boldsymbol{B}_0}{\partial t} \cdot d\boldsymbol{S}$ in plane S keeps unchanged also. Thus the induced electric field by $-\iint \frac{\partial \boldsymbol{B}_0}{\partial t} \cdot d\boldsymbol{S}$ does not vary with time, and $\boldsymbol{P}$ does not vary either. Therefore the polarization current in the medium is 0 during this period of time, i.e., the polarization of the toroid does not influence the magnetic field $\boldsymbol{B}_0$. Different kinds of the media have different $P$ values, and $P$ influences $E$ [13]. So the macro electric field $\boldsymbol{E}$ and $\oint_l \boldsymbol{E} \cdot d\boldsymbol{l}$ on the center line $l$ are different for different media. This means that $\oint_l \boldsymbol{E} \cdot d\boldsymbol{l}$ is not always equal to $-\iint \frac{\partial \boldsymbol{B}_0}{\partial t} \cdot d\boldsymbol{S}$. Hence Eqs. (2) and (5) are not correct in the media. They are correct only in vacuum.

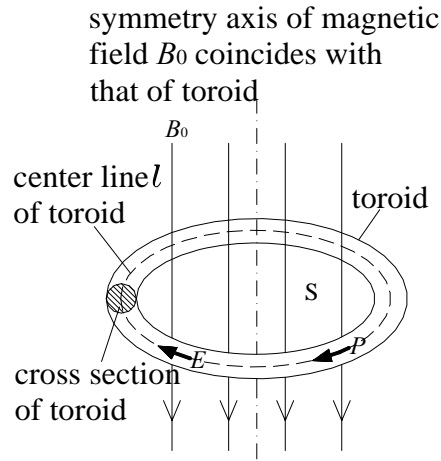

Figure 2 A varying magnetic field $\boldsymbol{B}_0$ passing through the toroid induces an electric field on the center line $l$ of the toroid, and this electric field produces a polarization $\boldsymbol{P}$ in the medium (the toroid). Since $\boldsymbol{P}$ reduces the electric field, the final macro electric field $\boldsymbol{E}$ on the center line $l$ of the toroid is different from medium to medium.

Usually, $\mu \approx \mu_0$ holds for most of the media, therefore the media have little influence on the exciting magnetic field $\boldsymbol{B}_0$. While different media often show quite different $\varepsilon$ values, and have quite different electric fields in the media under the same exciting electric field. So, if in Fig. 2 both the electric field and the magnetic field are in the media, it is also easy to illuminate that Eqs. (2) and (5) are not correct in the media.

In the next section, the existing Faraday's law of electromagnetic induction is modified so as to make it applicable in the media. With this modification a revised Maxwell's Equations in the media are obtained.

## 4. Maxwell's Equations in media

### 4.1 Without the induced magnetization in media

Suppose that there are the polarizations and the magnetic field magnetizations, but not the induced magnetization in the media under the EW. In the following text, $E$ and $B$ denote the electric field and the magnetic field in the media respectively, $E_0$ and $B_0$ denote the electric field and the magnetic field in vacuum respectively. Firstly, change Eq. (4) into

$$\nabla \times (B - \mu_0 M) = \mu_0 \varepsilon_0 \frac{\partial (E + P/\varepsilon_0)}{\partial t}, \qquad (6)$$

Eqs. (4) and (6) are reasonable in the media. They reflect the relationship between the change of the electric field ($E+P/\varepsilon_0$) and the curl of the magnetic ($B-\mu_0 M$)[14]. They are in agreement with Maxwell's electromagnetic theory: the change of an electric field produces a magnetic field, and the change of a magnetic field produces an electric field.

Similar to the relationship between ($E+P/\varepsilon_0$) and ($B-\mu_0 M$) in Eq. (6), according to "the change of a magnetic field produces an electric field", conversely Faraday's law of electromagnetic induction in the media should reflect the relationship between the change of ($B-\mu_0 M$) and the curl of ($E+P/\varepsilon_0$):

$$\nabla \times (E + P/\varepsilon_0) = -\frac{\partial (B - \mu_0 M)}{\partial t}. \qquad (7)$$

The difference between Eq. (7) and Eq. (2) is that Eq. (7) includes the polarization term and the magnetic-field magnetization term. Equation (7) shows that [12] $B$, $M$, $E$ and $P$ are all involved in the process of mutual transformation between the electric field and the magnetic field of the EW. Their energies should be regarded as the components of the energy of the EW. The propagation of the EW in the media is the process of mutual transformation between varying ($E+P/\varepsilon_0$) and varying ($B-\mu_0 M$). Ref. [12] calls $E_T=E+P/\varepsilon_0$ the "total electric field", and $B_T=B-\mu_0 M$ the "total magnetic field". The meanings of the total electric field $E_T$ can be explained with Eq. (6) as follows: Change of $E$ with time produces the current $\varepsilon_0 \partial E/\partial t$, and change of $P/\varepsilon_0$ with time produces the current $\partial P/\partial t$. These two currents independently exist, and each produces its own magnetic field. These two magnetic fields have the same phase, and together they form the magnetic field of the EW[15]. Since changes in both $E$ and $P/\varepsilon_0$ contribute to the formation of the magnetic field of the EW, $E_T=E+P/\varepsilon_0$ is called the total electric field. $B_T=B-\mu_0 M$ has a similar meanings as $E_T$. Note that the expressions of $E_T$ and $B_T$ given here are only applicable in the media of this paper.

Propagation of the EW in the media is the process of the mutual transformation between $E_T$ and $B_T$. Thus, in order to express the EWs in the media, we can simply replace $E_0$ and $B_0$ in Maxwell's Equations in vacuum with $E_T$ and $B_T$ and obtain Maxwell's Equations in media:

$$\nabla \cdot E_T = \rho / \varepsilon_0, \qquad (8)$$

$$\nabla \times E_T = -\frac{\partial B_T}{\partial t}, \qquad (9)$$

$$\nabla \cdot \boldsymbol{B}_\text{T} = 0, \tag{10}$$

$$\nabla \times \boldsymbol{B}_\text{T} = \mu_0 \varepsilon_0 \frac{\partial \boldsymbol{E}_\text{T}}{\partial t}. \tag{11}$$

Substituting $\boldsymbol{E}_\text{T} = \boldsymbol{E} + \boldsymbol{P}/\varepsilon_0$ and $\boldsymbol{B}_\text{T} = \boldsymbol{B} - \mu_0 \boldsymbol{M}$ into Eqs. (8) – (11) we have

$$\nabla \cdot (\boldsymbol{E} + \boldsymbol{P}/\varepsilon_0) = \rho/\varepsilon_0, \tag{12}$$

$$\nabla \times (\boldsymbol{E} + \boldsymbol{P}/\varepsilon_0) = -\frac{\partial (\boldsymbol{B} - \mu_0 \boldsymbol{M})}{\partial t}, \tag{13}$$

$$\nabla \cdot (\boldsymbol{B} - \mu_0 \boldsymbol{M}) = 0, \tag{14}$$

$$\nabla \times (\boldsymbol{B} - \mu_0 \boldsymbol{M}) = \mu_0 \varepsilon_0 \frac{\partial (\boldsymbol{E} + \boldsymbol{P}/\varepsilon_0)}{\partial t}. \tag{15}$$

$\boldsymbol{E}$, $\boldsymbol{B}$ and $\boldsymbol{M}$ are of curl fields [13], and there is no net charge in the media ($\rho$=0). So $\nabla\cdot\boldsymbol{E}=\nabla\cdot\boldsymbol{P}=\nabla\cdot\boldsymbol{B}=\nabla\cdot\boldsymbol{M}=0$ [13]. Therefore Eqs. (12) – (15) can be changed into

$$\nabla \cdot \boldsymbol{E} = 0, \tag{16}$$

$$\nabla \times \boldsymbol{E} = -\frac{\partial \boldsymbol{B}}{\partial t} + \mu_0 \frac{\partial \boldsymbol{M}}{\partial t} - \nabla \times \boldsymbol{P}/\varepsilon_0, \tag{17}$$

$$\nabla \cdot \boldsymbol{B} = 0, \tag{18}$$

$$\nabla \times \boldsymbol{B} = \mu_0 \varepsilon_0 \frac{\partial \boldsymbol{E}}{\partial t} + \mu_0 \frac{\partial \boldsymbol{P}}{\partial t} + \mu_0 \nabla \times \boldsymbol{M}. \tag{19}$$

Comparing Eqs. (16) – (19) with Eqs. (1) – (4), we can see that Eqs. (16), (18) and (19) are identical to Eqs. (1), (3) and (4) respectively, except for Eq. (17), which is derived from Eq. (13). Eq. (17) and Eq. (13) show that the change of $\boldsymbol{B}$ and the change of $-\mu_0\boldsymbol{M}$ together produce an electric filed, then this electric filed produces the polarization $\boldsymbol{P}$ and the compositive field $\boldsymbol{E}$. Eq. (17) and Eq. (13) are not only applicable in vacuum but also in the media. The integral form of Eq. (13) is

$$\oint (\boldsymbol{E} + \boldsymbol{P}/\varepsilon_0) \cdot \mathrm{d}\boldsymbol{l} = -\iint \frac{\partial (\boldsymbol{B} - \mu_0 \boldsymbol{M})}{\partial t} \cdot \mathrm{d}\boldsymbol{S}. \tag{20}$$

Eq. (20) accords with the experiment in Fig. 2: In the center $l$, $\boldsymbol{E}+\boldsymbol{P}/\varepsilon_0$ is equal to $\boldsymbol{E}_0$ [16] which is the field when the toroid does not exist, i.e. $\boldsymbol{E}_0$ is the field in vacuum. $\boldsymbol{B}-\mu_0\boldsymbol{M}$ is equal to in $\boldsymbol{B}_0$ which is the magnetic field in vacuum (because $\boldsymbol{M}$=0). So Eq. (20) becomes $\oint_l \boldsymbol{E}_0 \cdot \mathrm{d}\boldsymbol{l} = -\iint \frac{\partial \boldsymbol{B}_0}{\partial t} \cdot d\boldsymbol{S}$. Obviously, it is correct.

In a word, Faraday's law of electromagnetic induction in the media should reflect the relationship between the field ($\boldsymbol{B}-\mu_0\boldsymbol{M}$) and the field ($\boldsymbol{E}+\boldsymbol{P}/\varepsilon_0$). It can write as Eq. (7), Eq. (9), Eq. (17) or Eq. (20). These equations have the same meanings, just in different forms.

### 4.2 With the induced magnetizations in media

In this section the following situation will be considered: Under the EW, there are not only the magnetic field magnetization and the polarization, but also the induced magnetization in the media. As mentioned in Section 2.2, the magnetic field of the induced magnetization has a different phase with the magnetic field of the EW, so the magnetic field of the induced magnetization is not a component of the magnetic field of the EW itself. It is well known that, when describing the EW, the static fields should not be considered in Maxwell's Equations because they do not belong to the fields of the EW itself. For the similar reason, we believe that the field and the current of the induced magnetization should not be included in Maxwell's Equations when describing the EW. Thus Maxwell's Equations in the media in this section are identical to those obtained in the last section.

In short, whether in vacuum or in the media, and regardless whether there is the induced magnetization in the media, Maxwell's Equations can write as Eqs. (8)-(11) or Eqs. (16)-(19) when they are used to describe the EWs.

## 5. Solution of Maxwell's equations

Inspired by the solution of Maxwell's equations in vacuum which can be found in many related books, we obtain a simple-harmonic-wave solution to Eqs. (8) - (11):

$$E_T = E_{Tm} \cos(\omega t - \frac{k}{n} z + \phi)$$
$$B_T = B_{Tm} \cos(\omega t - \frac{k}{n} z + \phi)$$
, (21)

where $E_{Tm}$ and $B_{Tm}$ are the amplitudes of $E_T$ and $B_T$ respectively, $\omega$ is the angle speed, $k$ is the wave number, $n$ is the reference index, $\phi$ is the initial phase. Propagation direction of the EW is along the z direction of the rectangular coordinate system, and $E_T$ and $B_T$ point to the x direction and the y direction respectively. $n$ is also the ratio of the energy of the EW in vacuum over that in the media, or the ratio of the volume of the EW in vacuum over that in the media[1]. Eq. (21) is the solution of the EW not only in vacuum but also in the media. From Eqs. (8)-(11) and the solution Eq. (21) we have

$$\frac{k}{n} = \omega \sqrt{\varepsilon_0 \mu_0} = \frac{\omega}{c}$$ , (22)

or

$$k = \frac{\omega}{c/n} = \frac{\omega}{u}$$ , (23)

where $u=c/n$, $c=1/(\varepsilon_0\mu_0)^{1/2}$. $u$ is propagation velocity of the EW. If there is no other known condition, $u$ or $n$ value cannot be obtained from Eqs. (8) - (11) and the solution. $u$ or $n$ is the undetermined parameter of Eq. (21), just like $\omega$ and $\phi$. $u$ or $n$ can be obtained from measurements or calculations. It can be seen that $n$ is not directly related to $\varepsilon_r$ and $\mu_r$. In other words, refraction of the EWs in the media is not directly related to the polarization and the magnetic field magnetization. The polarization and the magnetic field magnetization influence $n$ value indirectly [1]. From Eqs. (8)-(11) and the solution Eq. (21) we also obtain

$$E_T = cB_T$$ , (24)

In the uniform and infinite media, which is the precondition of this paper, $D=\varepsilon_0 E_0$ [17], where $D=\varepsilon_0 E+P$ and $D$ is electric displacement [13]. So

$$E_T = E+P/\varepsilon_0 = D/\varepsilon_0 = E_0 ,  \quad (25)$$

and

$$B_T = B_0 . \quad (26)$$

Therefore $E_{Tm}=E_{0m}$ and $B_{Tm}=B_{0m}$, where $E_{0m}$ and $B_{0m}$ are respectively the amplitudes of the electric field and the magnetic field of the EW in vacuum. Usually, EWs' intensities decline due to reflection, absorption and scattering as they pass the media. At the Introduction it is stated that we only consider the refraction parts of EWs, and do not consider the lost parts of EWs due to reflection, absorption and scattering etc. So Eq (26) only deals with the refraction parts of EWs. Hence in both vacuum and the media, regardless whether there is the induced magnetization in the media, another form of solution of Eqs. (8)-(11) is

$$E_T = E_{0m} \cos(\omega t - \frac{k}{n} z + \phi)$$
$$B_T = B_{0m} \cos(\omega t - \frac{k}{n} z + \phi) \quad , \quad (27)$$

In Eq. (21) or Eq. (27) $n$ value in vacuum is different from that in the media.

## 6. Conclusion

The existing Faraday's law of electromagnetic induction $\oint E \cdot dl = -\iint \frac{\partial B}{\partial t} \cdot dS$ is not reasonable in media because it does not include the polarization item and the magnetic field magnetization item. In insulation, homogeneous, isotropic and infinity media, regardless of whether there is the induced magnetization in the media, Maxwell's equations for describing the EWs are

$$\nabla \cdot E_T = 0 ,$$

$$\nabla \times E_T = -\frac{\partial B_T}{\partial t} ,$$

$$\nabla \cdot B_T = 0 ,$$

$$\nabla \times B_T = \mu_0 \varepsilon_0 \frac{\partial E_T}{\partial t} .$$

These equations are also applicable in vacuum. Propagation of the EWs in the media is the process of mutual transformation between varying $(E+P/\varepsilon_0)$ and varying $(B-\mu_0 M)$. A simple-harmonic-wave solution to the above equations is

$$E_T = E_{0m} \cos(\omega t - \frac{k}{n} z + \phi)$$
$$B_T = B_{0m} \cos(\omega t - \frac{k}{n} z + \phi)$$

$n$ is the undetermined parameter of the above solution, just like $\omega$ and $\phi$. $n$ value in vacuum is different from that in the media. Refraction of the EWs in the media is not directly related to the polarization and the magnetic field magnetization.